\documentclass[tighten,twocolumn]{aastex63}

\usepackage{graphicx}
\usepackage{ulem}
\usepackage{amsmath}
\usepackage{wrapfig}
\usepackage[thinlines]{easytable}
\usepackage{tikz}
\usepackage{hyperref}
\usepackage{braket}

\graphicspath{{./}{figs/}}

\begin{document}

\title{The Impact of High-Mass Clusters on the Binary Black Hole Merger Rate over Cosmic Time}

\correspondingauthor{Aidan Mai}
\email{aimai@unc.edu}

\author[0009-0003-8963-7787]{Aidan Mai}
\affiliation{Department of Physics and Astronomy, University of North Carolina at Chapel Hill; Chapel Hill, NC 27599, USA}
\affiliation{Department of Astronomy \& Astrophysics, University of California, San Diego; La Jolla, CA 92093, USA}

\author[0000-0002-4086-3180]{Kyle Kremer}
\affiliation{Department of Astronomy \& Astrophysics, University of California, San Diego; La Jolla, CA 92093, USA}

\author[0000-0003-4412-2176]{Fulya K{\i}ro\u{g}lu}
\affiliation{Center for Interdisciplinary Exploration \& Research in Astrophysics (CIERA) and Department of Physics \& Astronomy \\ Northwestern University, Evanston, IL 60208, USA}
\affil{NSF-Simons AI Institute for the Sky (SkAI), 172 E. Chestnut St., Chicago, IL 60611, USA}

\begin{abstract}

We present a cosmological calculation of the binary black hole merger rate from dense star clusters, using a \texttt{Cluster Monte Carlo} (\texttt{CMC}) model grid spanning the full observed parameter space of cluster mass, metallicity, and virial radius. For the first time, we incorporate $N=10^{7}$-particle \texttt{CMC} models into this calculation, including a new solar-metallicity model, extending the \texttt{CMC Cluster Catalog} to the most massive clusters observed in the local Universe. Using an analytic cluster formation model which convolves the cluster formation rate with delay-time distributions from our model suite, we find that the local cluster BBH merger rate is dominated by low-mass ($M_1 \lesssim 20\,M_\odot$) mergers originating from recently formed, high-metallicity clusters. While high-mass clusters are the primary sites of hierarchical (second-generation and higher) mergers, their contribution to the overall merger rate is suppressed by the exponential cutoff of the cluster initial mass function. The hierarchical merger rate is therefore highly sensitive to the exact value of the Schechter mass, $M_S$, suggesting that gravitational wave constraints on the hierarchical merger fraction can independently constrain the high-mass end of the cluster initial mass function. Our cluster formation model reproduces the observed present-day GC number density without fine tuning, and we find $M_S = 10^7\,M_\odot$ provides the best match to the observed Virgo GC mass distribution. For this value, $27\%$ of local dynamical mergers are hierarchical, compared to $18\%$ for the canonical value of $M_S = 10^{6.26}\,M_\odot$. Our merger weights, which we make publicly available, can be used for direct comparison with LVK distributions and other formation models.

\vspace{1cm}
\end{abstract}

\section{Introduction}
\label{sec:intro}

The growing catalog of gravitational wave (GW) sources detected by the LIGO--Virgo--KAGRA (LVK) collaboration over the past decade has revealed a diverse population of merging binary black holes \citep[BBHs;][]{LVK2026_GWTC5}. Several features within the catalog's black hole (BH) mass and spin distributions suggest that no single evolutionary pathway is likely to account for the entire observed population \citep[e.g.,][]{Zevin2021,LVK2026_GWTC5}. Proposed formation channels include isolated binary evolution in galactic fields \citep[e.g.,][]{Belczynski2016_first,Marchant2016,Bavera2021_impact}, dynamical assembly in dense stellar systems \citep[e.g.,][]{PortegiesZwartMcMillan2000,Rodriguez2016mergers,AntoniniGieles2020}, and mergers in disks of active galactic nuclei \citep[][]{Stone2017,McKernan2018,Tagawa2020}. Determining the relative contribution of these channels to the overall merger rate remains a central objective in gravitational wave astrophysics \citep[e.g.,][]{MandelFarmer2022,MandelBroekgaarden2022}.

Dynamical assembly from dense star clusters in particular is now viewed as a canonical formation scenario. Through repeated dynamical encounters, initially unrelated BHs can form binaries and undergo repeated hardening until they merge, either inside their host cluster or after dynamical ejection \citep[e.g.][]{Sigurdsson1993,PortegiesZwartMcMillan2000,Morscher2015,Rodriguez2016mergers,Askar2017,Banerjee2017,Rodriguez2018, Samsing2018,Fragione2019,DiCarlo2019,Kremer2020modeling}. Dynamical BBH systems are characterized by spin-orbit misalignment \citep[e.g.][]{Rodriguez2016_spin, Zevin2021,Payne2024}, high component masses above the pair-instability gap \citep[through repeated ``hierarchical'' mergers; e.g.][]{MillerHamilton2002, GerosaBerti2017,Rodriguez2019, Antonini2019,FragioneRasio2023,Kiroglu2025c,Mai2026colossus}, and/or high merger eccentricity \citep[e.g.][]{Samsing2014,Zevin2021_eccentricity}, all of which can be difficult to obtain through standard isolated binary evolution. 
Indeed, several individual GW events have properties consistent with a dynamical or hierarchical origin, including GW190521 \citep[e.g.,][]{LIGO2020,RomeroShaw2020}, GW190412 \citep[e.g.,][]{GW190412,Rodriguez2020_SSC}, GW231123 \citep[e.g.][]{GW231123,Mai2026colossus}, and the recent asymmetric, high-spin mergers GW241011 and GW241110 \citep[e.g.,][]{GW241011,Agrawal2026metallicity}.


The term ``dense star cluster,'' however, encompasses more than the canonical picture of old, metal-poor globular clusters (GCs). Massive bound clusters span a wide range of ages, metallicites, masses, and galactic environments \citep[for review, see][]{PortegiesZwart2010}, and recent studies have shown that the full ranges of both cluster mass and metallicity can contribute substantially to the local BBH merger rate, especially recently-formed, high-metallicity clusters \citep[e.g.][]{Choksi2019,Rodriguez2020_SSC,AntoniniGieles2020,Mapelli2021,Bruel2024A,Ye2026,Agrawal2026metallicity}. In high-metallicity clusters, line-driven winds from massive stars generally become stronger, enhancing mass loss prior to core collapse and reducing the masses of the resulting stellar remnants \citep[e.g.][]{Vink2001,Spera2015}. Stellar evolution calculations therefore typically predict lower average and maximum BH masses at higher metallicity \citep[e.g.,][]{Spera2015, Vink2024}. This relationship is not uniquely determined by stellar winds: binary mass transfer and stripping, rotation and internal mixing, core-collapse and fallback physics, and pulsational pair-instability can all modify the BH mass function \citep[e.g.][]{Fryer2012,Farmer2019,Schneider2023,Gallegos-Garcia2026}. Direct observational evidence for a BH mass--metallicity relation also remains sparse and complicated by strong, channel-dependent selection effects \citep[][]{Liotine2023,Gaia2024,Balbinot2024}. Nevertheless, the broad theoretical expectation is that lower-metallicity populations generally shift the first-generation BH mass spectrum toward higher masses, whereas near-solar-metallicity populations preferentially produce lower-mass BHs. Consequently, old, metal-poor clusters preferentially contribute higher-mass BBH mergers, while metal-rich clusters can supply the low-mass mergers that dominate the local BBH population \citep[e.g.,][]{Chatterjee2017_highZ,DiCarlo2020,Mapelli2021,Banerjee2021,Ye2026,Agrawal2026metallicity}.

The contribution of high-metallicity clusters to the local merger rate is corroborated by events like GW241110 and GW241011, whose large spin-orbit misalignment and low component masses suggest a dynamical origin from a metal-rich cluster \citep[][]{GW241011}, and emphasizes that hierarchical formation need not be exclusively associated with high-mass pair-instability-gap events from metal-poor GCs. Therefore, a cosmological calculation of dynamical mergers must incorporate the full range of observed cluster metallicity, whose redshift dependence couples the redshift evolution of BBH merger masses and rates to the cosmic formation history of dense clusters \citep[e.g.,][]{RodriguezLoeb2018,Choksi2019,Ye2026,OConnor2026_LVK}.

Cluster mass introduces a similarly important dependence. More massive clusters not only contain more BHs overall, but they are also more efficient at merging their BHs; recent studies show the number of BBH mergers scales roughly as $N_{\rm BBH} \propto M_{\rm cl}^{1.4}$ \citep{Hong2018,AntoniniGieles2020,Mai2026colossus}, steeper than the linear scaling one may guess from cluster mass alone. Their higher escape velocities also aid in retention of merger remnants receiving gravitational wave recoil kicks, enabling the formation of second- and higher-generation systems with large component masses and spins \citep[e.g.,][]{Rodriguez2019,GerosaBerti2019}. The yield of hierarchical mergers therefore rises more steeply with cluster mass than first-generation mergers \citep[e.g.,][]{Mai2026colossus}. The competition between the steep cluster mass function and the superlinear merger yield distributes first-generation mergers across a broad range of cluster masses, whereas hierarchical mergers are much more strongly concentrated in the high-mass tail \citep[e.g.,][]{AntoniniGieles2020,Mai2026colossus}.

\begin{figure*}
    \includegraphics[width=\textwidth]{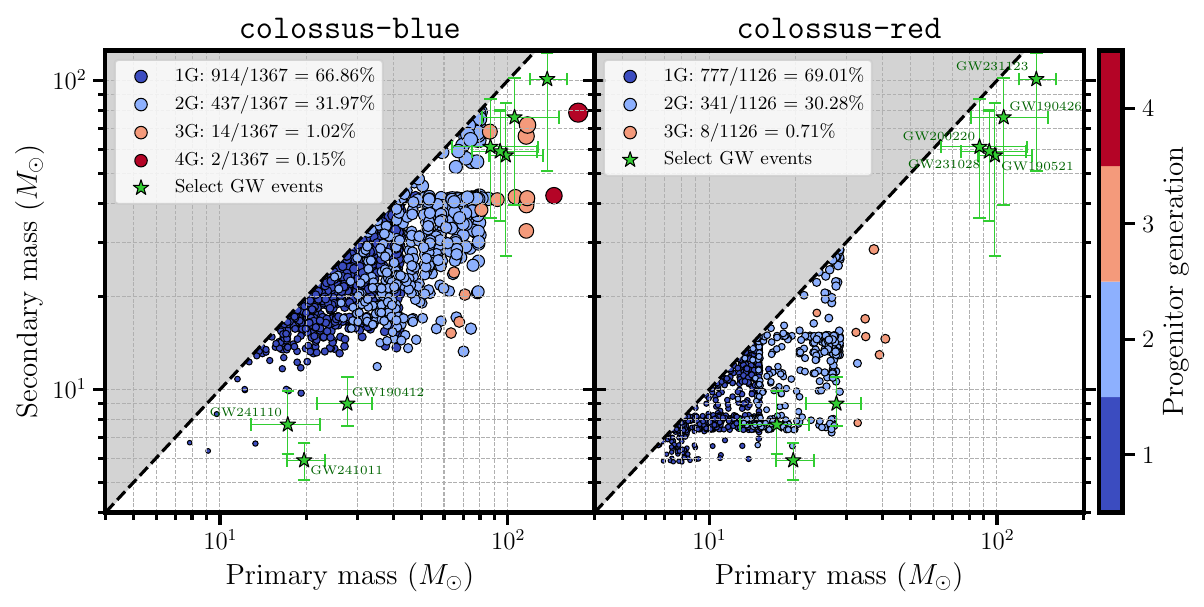}
    \caption{Primary vs. secondary mass for all BBH mergers in \texttt{colossus-blue} and \texttt{colossus-red}, which are identical $N=10^7$ models except for differing metallicity: $Z/Z_\odot = 0.1$ and $1.0$, respectively. Each merger is colored by hierarchical generation. While both models are roughly similar in terms of total number of mergers and 2G+ fraction, the bottom-heavy high-metallicity BH mass function of \texttt{colossus-red} results in much lower component masses than \texttt{colossus-blue}, producing no pair-instability-gap events even at third generation. We also show observed component masses for eight select LVK events (five high-mass and three low-mass) that have been associated with possible hierarchical merger origin. Clusters of both high- and low-metallicity are needed to span the full observed primary-secondary parameter space.}
    \label{fig:m1m2-crb}
\end{figure*}

However, the exact shape of this high-mass tail remains uncertain. The initial cluster mass function is commonly described by a \citet{Schechter1976}-like power law modified by a high-mass exponential truncation: \(dN/dM\propto M^{-2}\exp(-M/M_S)\) \citep[e.g.,][]{Gieles2006,Bastian2008,Larsen2009,Bastian2012,Johnson2017,Messa2018a}. Observations provide evidence that both the characteristic truncation mass $M_S$ and the shape of the high-mass tail vary with galactic environment \citep[e.g.,][]{Bastian2012,Adamo2015,Johnson2017,Messa2018b}. Theoretical models show the maximum cluster mass is dependent upon gas surface density, pressure, shear, and feedback within the host galaxy at cluster formation \citep[e.g.,][]{Kruijssen2014,ReinaCampos2017,Pfeffer2018,Grudic2023}. Furthermore, the universality of an exponential truncation remains debated, with some cluster populations consistent with an untruncated power law over the observed mass range \citep[e.g.,][]{Whitmore2010,Chandar2011,Mok2019}. The mapping between the initial and present-day cluster mass functions is further modified by stellar-evolution mass loss, two-body evaporation, tidal shocks, dynamical friction, and even dissolution driven by retained BH populations \citep[e.g.,][]{GnedinOstriker1997,FallZhang2001,BaumgardtMakino2003,Gieles2009,Kruijssen2015,Pfeffer2018,GielesGnedin2023}.

In this study, we extend recent \texttt{CMC}-based cosmological BBH merger-rate calculations by incorporating the most massive \texttt{CMC} models available, including the low-metallicity $N=10^7$ \texttt{colossus} simulation presented in \citet{Mai2026colossus}, and a new solar-metallicity $N=10^7$ counterpart that we dub ``\texttt{colossus-red}''. The latter is representative of both the most massive ``red'' globular clusters observed in nearby massive elliptical galaxies \citep[e.g.,][]{Peng2006,Jordan2007,Harris2009,Strader2011}, and of the young massive super star clusters observed in the local Universe \citep[e.g.,][]{WhitmoreSchwiezer1995,Bastian2006,McCradyGraham2007,PortegiesZwart2010}. This is especially important for hierarchical mergers, whose yield rises steeply with cluster mass. 

To highlight the role of metallicity in BBH merger masses and hierarchical merger formation, we show in Figure~\ref{fig:m1m2-crb} primary versus secondary mass for all BBH mergers in \texttt{colossus-red} and the original \texttt{colossus} model (now named \texttt{colossus-blue}). Both models have similar total BBH merger yields ($1367$ and $1126$ for \texttt{blue} and \texttt{red}, respectively) and hierarchical merger fractions ($33\%$ and $31\%$), but \texttt{colossus-red} has much lower average and maximum primary masses as a result of its high-metallicity BH mass function. We also show eight select GW events previously discussed in the literature as being of possible dynamical origin: three low-mass events with high spin-orbit misalignment \citep{GW241011,GW190412}, and the five current highest-mass GW events in the cumulative GWTC-5.0 catalog \citep{GW231123,GWTC21,GW190521}. Both high and low cluster metallicites are required to span the observed parameter space of possible dynamical events.

We combine these models with a cosmological cluster-formation framework to determine how the total and hierarchical BBH merger rates, mass distributions, and redshift evolution depend on the high-mass cutoff of the initial cluster mass function, as well as assumptions on initial cluster virial radii. This emphasis on the extreme high-mass tail complements recent \texttt{CMC} studies focused primarily on metallicity and cluster age \citep{Ye2026,Agrawal2026metallicity} or on realistic primordial-binary populations \citep{OConnor2026_LVK}. Furthermore, as many of these massive clusters may potentially host massive black holes in their centers \citep{Haberle2024}, this work sets the stage for discussion of the potential role of intermediate-mass BHs on the formation of gravitational wave sources (Wu et al. 2026, in prep).

This paper is organized as follows. In Section~\ref{sec:cmc}, we describe the \texttt{CMC} code and the properties of our model grid. In Section~\ref{sec:integration}, we introduce our analytic cluster formation model and discuss the redshift evolution of its resultant BBH merger population. In Section~\ref{sec:virgo}, we compare the present-day properties of observed Virgo GCs with our formation model, and vary our initial cluster mass and virial radius functions to better match the Virgo population. In Section~\ref{sec:bbh}, we discuss how these parameter variations affect the mass and spin distributions of our model's BBH mergers. We discuss and conclude in Section~\ref{sec:conclusion}.

\section{Modeling Stellar Clusters with CMC}
\label{sec:cmc}

To study the long-term dynamical evolution of dense star clusters and the BBHs they form, we use the \texttt{Cluster Monte Carlo} code \texttt{CMC} \citep{Rodriguez2022}. Specifically, we use the $148$ models published as part of the \texttt{CMC Cluster Catalog} \citep{Kremer2020modeling}, supplemented by the $26$ new models from \citet{Agrawal2026metallicity} that refine the coverage of intermediate metallicities. To extend model coverage to the most massive clusters observed, we also include two $N=10^7$-star simulations with metallicity $\{0.1,1\}\,Z_\odot$. Following \citet{Mai2026colossus}, we refer to these two massive models as \texttt{colossus-blue} and \texttt{colossus-red}, respectively. In total, our suite of $176$ independent models covers a range in $N\in \{2,\,4,\,8,\,16,\,32,\,100\}\times10^5$, virial radius $r_v\in \{0.5,\,1,\,2,\,4\}\,$pc, galactocentric position $R_{\rm gc}\in\{2,\,8,\,20\}\,$kpc (assuming a Milky-Way-like galactic potential), and metallicity $[\rm{Fe/H}]\in\{-2,-1.5,-1,-0.75,-0.5,-0.25,0,0.1\}$ that collectively spans the parameters space of massive star clusters, both young and old, that are observed in the local Universe.

For detailed discussion of \texttt{CMC} methods, we refer the reader to \citet{Rodriguez2022} and for discussion of the specific model assumptions made in these models, to \citet{Kremer2020modeling,Agrawal2026metallicity}. We summarize here a few key details. We adopt a \citet{Kroupa2001} initial stellar mass function in the mass range $0.08-150\,M_\odot$. All models adopt a primordial binary fraction of $5\%$, uniform across all stellar masses. Initial binary orbital periods are drawn from a distribution flat in log-scale from near contact to the local hard-soft boundary, and initial eccentricities are drawn from a thermal distribution \citep{Heggie1975}. All simulations adopt the ``rapid model'' from \citet{Fryer2012} for stellar remnant formation. Neutron star natal kicks are drawn from a Maxwellian with dispersion $\sigma=265\,$km/s \citet{Hobbs2005}. Black hole natal kicks are drawn from the same Maxwellian but scaled down based on the amount of mass fallback prescribed by the \citet{Fryer2012} model. The (pulsational) pair instability prescription follows \citet{Belczynski2016}. Following \citet{FullerMa2019spin}, we assume all first generation black holes born via stellar collapse are formed with zero spins. Following black hole mergers, new spins, masses, and gravitational wave recoil kicks are computed following \citet{Rodriguez2018} and references therein.

\section{Cosmic Integration}
\label{sec:integration}

Combining the model suites of \cite{Kremer2020modeling} and \cite{Agrawal2026metallicity}, along with both \texttt{colossus} models, we obtain a model grid that fully spans the complete $M_{\rm cl}$-$r_v$-$Z$ parameter space of observed dense clusters. We can now compute the cosmological BBH merger rate from clusters of all parameters across redshift. We aim to approximate the convolution
\begin{equation}
\label{eq:convolution}
\mathcal{R}_{\rm BBH}(t) = \int_0^t \mathcal{R}_{\rm GC}(t')P(t-t')dt'
\end{equation}
where $\mathcal{R}_{\rm GC}(t')$ is the GC formation rate density at time $t'$, and $P(t-t')$ is the BBH delay-time distribution, encompassing the probability that a BBH merger occurs at time $t-t'$ after cluster formation.

\subsection{Cluster Formation Model}
\label{sec:formation}
For $\mathcal{R}_{\rm GC}$, we adopt the same cluster formation model as \cite{Ye2024redshift}. We first assume a Schechter cluster mass function
\begin{equation}
p(M_{\rm GC}) \propto \left(\frac{M_{\rm GC}}{M_S}\right)^{\beta_S}\exp\left(-\frac{M_{\rm GC}}{M_S}\right)
\label{eq:schechter}
\end{equation}
where the Schechter mass is $M_S = 10^{6.26}M_\odot$ and $\beta_S = -2$, from $M_{\rm cl} \in [10^5~M_\odot,10^7~M_\odot]$ \citep[compared to a maximum of $10^6~M_\odot$ in][]{Ye2024redshift}.

\begin{figure*}
    \includegraphics[width=\textwidth]{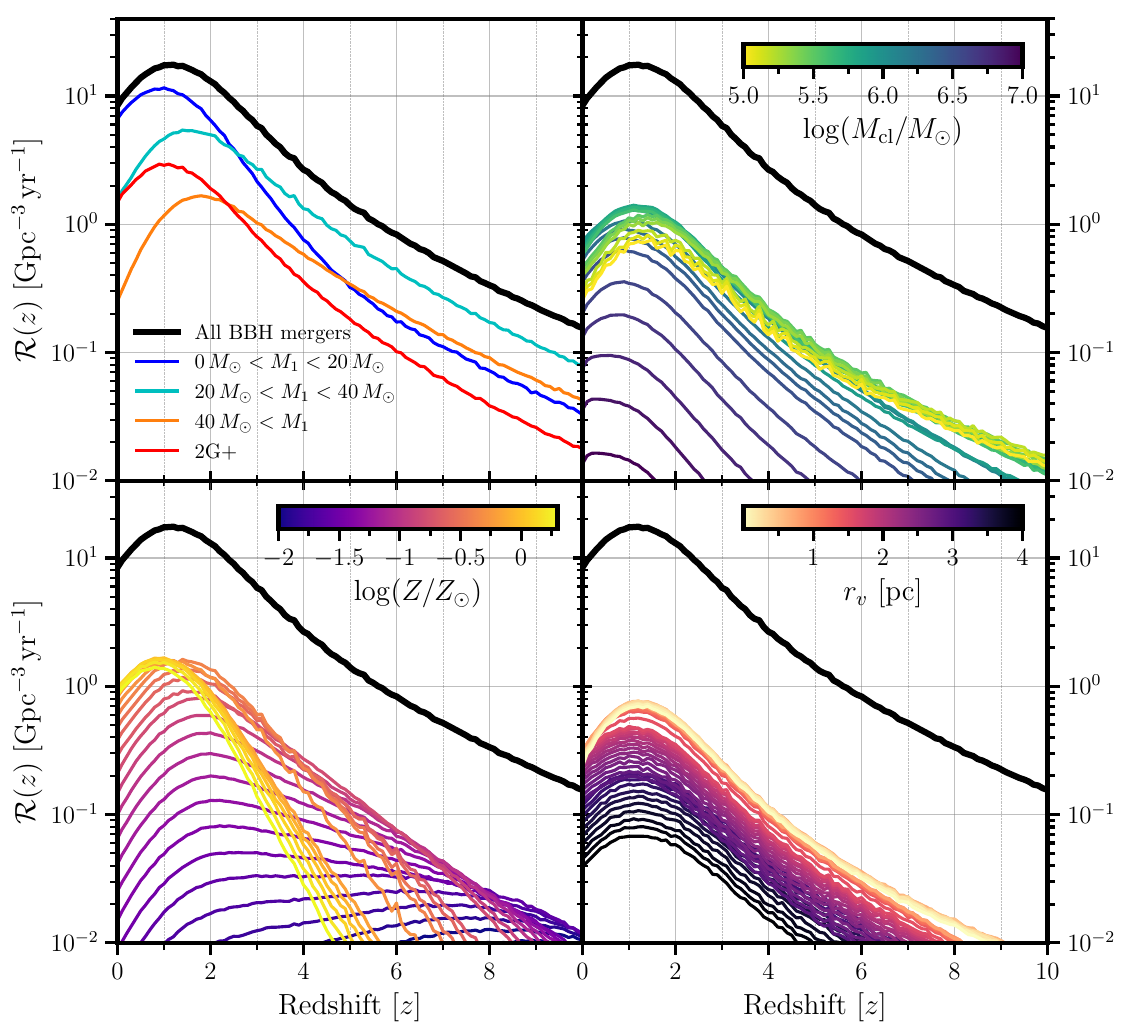}
    \caption{Contributions of different cluster and merger types to the dynamical BBH merger rate over redshift, obtained using the cosmic integration process described in Section~\ref{sec:integration}. Within our formation model, low-mass ($M_1 < 20\,M_\odot$) mergers from young, high-metallicity clusters dominate the merger rate at low redshifts ($z < 2$). The contribution of high-mass clusters beyond the assumed Schechter mass, $M_S = 10^{6.26}\,M_\odot$, is suppressed due to the exponential term in Equation~\ref{eq:schechter}.}
    \label{fig:rate}
\end{figure*}

We adopt a Gaussian distribution in cluster virial radii from $r_v \in [0~\rm pc, 4~pc]$:
\begin{equation}
p(r_v) \propto \exp\left(-\frac{(r_v - \mu_r)^2}{2\sigma_r^2}\right)
\label{eq:rvpdf}
\end{equation}
where $\mu_r = \sigma_r = 2~\rm pc$ \citep{Ye2024redshift}. We note that our BBH scaling relation with $r_v$ is an exponential function rather than a power law (discussed later in Equation~\ref{eq:wscale}), so in the context of BBH rate calculations, merger rates are indeed well-behaved as $r_v \rightarrow 0$.

We assume a lognormal metallicity distribution from $\log (Z/Z_\odot) \in [-6, 0.3]$:
\begin{equation}
p(Z) \propto \exp\left(-\frac{(\log Z - \mu_Z)^2}{2\sigma_Z^2}\right)
\label{eq:zpdf}
\end{equation}
where $\sigma_Z = 0.5$, and $\mu_Z$ is given by the following redshift-dependent equation from \cite{MadauFragos2017}:
\begin{equation}
\mu_Z = \log\braket{Z/Z_\odot} = 0.153 - 0.074z^{1.34}.
\label{eq:muz}
\end{equation}

Next, we adopt the star formation rate density also given in \cite{MadauFragos2017}:
\begin{equation}
\psi(z) = 0.01 \frac{(1+z)^{2.6}}{1 + [(1+z)/3.2]^{6.2}}~[M_\odot\rm{yr^{-1} Mpc^{-3}}]
\label{eq:sfrd}
\end{equation}
and define $f_{\rm cl}$ to be
\begin{equation}
f_{\rm cl} = \frac{\int_{10^5 M_\odot}^{10^7 M_\odot} Mp(M)dM}{\int_{10^2 M_\odot}^{10^7 M_\odot} Mp(M)dM} = 0.258,
\end{equation}
denoting the fraction of star formation mass used for massive star cluster formation, where $p(M)$ is the Schechter function given in Equation~\ref{eq:schechter}. We also adopt $f_{\rm bound} = 1/20$ to denote the fraction of embedded clusters which remain bound after star formation \citep{LadaLada2003}, and denote $f_{\rm disrp} = 1/5$ as the fraction of bound clusters which survive to present day \citep[e.g.][]{Ye2024redshift,AntoniniGieles2020mergers}. We then obtain
\begin{equation}
\mathcal{R}_{\rm GC}(z) = \frac{f_{\rm cl}f_{\rm bound}f_{\rm disrp}\psi(z)}{\braket{M_{\rm GC}}}~[\rm{yr^{-1}Mpc^{-3}}]
\label{eq:rgc}
\end{equation}
where $\braket{M_{\rm GC}}=10^{5.46}~M_\odot$ is the mean GC mass between $10^5~M_\odot$ and $10^7~M_\odot$, weighted by $p(M)$.

Next, we numerically evaluate the convolution by sampling the BBH delay-time distributions of our \texttt{CMC} models. We discretize the $M_{\rm cl}$-$r_v$-$Z$ parameter space into a uniform grid, with $\log (M_{\rm cl}/M_\odot) \in [5.0,7.0]$, $r_v / {\rm pc} \in [0.0,4.0]$, and $\log (Z/Z_\odot) \in [-6.0, 0.3]$, and a step size of $0.1$ in all dimensions.

We aim to map each $M_{\rm cl}$-$r_v$-$Z$ point to its closest \texttt{CMC} model counterpart, in order to sample its delay-time distribution. We first sort the points into metallicity bins, then map them to their nearest \texttt{CMC} neighbor within the same bin in $M_{\rm cl}$-$r_v$ space. \texttt{CMC} model metallicities are taken from $\log(Z/Z_\odot) \in \{-2, -1.5,-1,-0.75,-0.5,-0.25,0,0.1\}$, and we assign each $M_{\rm cl}$-$r_v$-$Z$ point to draw from the two closest \texttt{CMC} metallicities in $\log(Z/Z_\odot)$ space. For instance, a point with $\log(Z/Z_\odot)=-1.2$ lies between \texttt{CMC} metallicities of $-1$ and $-1.5$, so it would draw equally from \texttt{CMC} models with $\log(Z_{\texttt{CMC}}/Z_\odot) \in \{-1.5,-1\}$. For points with $\log(M_{\rm cl}/M_\odot) > 6$, due to sparser metallicity coverage at high mass, we simply sort into binary high- and low-metallicity groups, with $\log(Z_{\texttt{CMC}}/Z_\odot) \in \{-2,-1\}$ or $\{0\}$.

Next, within each metallicity group, we map each $M_{\rm cl}$-$r_v$-$Z$ point to its closest \texttt{CMC} model counterpart using the method described in \cite{Mai2026colossus}. We define the following distance function in $M_{\rm cl}$-$r_v$ space:
\begin{equation}
    \label{eq:distance}
    \begin{aligned}
        d^2  & = (C \cdot \Delta \log M)^2 + (\Delta r_v)^2 \\
           & = \left(C\log \frac{M_{\texttt{CMC}}}{M_\text{cl}}\right)^2 + \left(r_v^{\texttt{CMC}} - r_v\right)^2
    \end{aligned}
\end{equation}
where $C=10$ is a constant that accounts for the differing scales between $\log M$ and $r_v$. We map each point to the \texttt{CMC} model within the same metallicity bin that minimizes $d$. We choose randomly between models with degenerate $M_{\rm cl}$ and $r_v$ within their respective metallicity bins.

We weight each point by the likelihood of its formation properties, given in Equations~\ref{eq:schechter}, \ref{eq:rvpdf}, and \ref{eq:zpdf}:
\begin{equation}
    \label{eq:wform}
    w_{\rm form} = p(M_{\rm cl})p(r_v)p(Z|z_f).
\end{equation}
Furthermore, to approximate continuous dependence on cluster properties, we apply a scaling correction to each point based on its deviation from the nearest \texttt{CMC} model, using the scaling relations derived in \cite{Mai2026colossus}:
\begin{equation}
    \label{eq:wscale}
    w_{\rm scale} = \left(\frac{M_{\rm GC}}{M_{\texttt{CMC}}}\right)^{\beta}\exp\left[-\alpha(r_v - r_{v}^{\texttt{CMC}})\right]
\end{equation}
where $\beta = 1.4$ and $\alpha = 0.5$ for 1G mergers, and $\beta = 2.0$ and $\alpha = 1.0$ for 2G+ mergers. Note that when scaling with the \emph{present-day} property $r_h$ as in \cite{Mai2026colossus}, $\alpha$ is dependent on $Z$, since the metallicity-dependent BH mass function imprints itself upon present-day properties through various dynamical processes such as black hole burning, mass segregation, and eventual core collapse \citep[e.g.][]{Agrawal2026metallicity}. On the other hand, \emph{initial} cluster properties like $r_v$ are free of this metallicity dependence, and since we consider a wide range of metallicities in this analysis, we choose to scale off of $r_v$ instead. Indeed, we find both $\alpha$ scalings to approximately hold for both low and high metallicity groups.

Combining Equations~\ref{eq:rgc},~\ref{eq:wform},~and~\ref{eq:wscale}, for each $M_{\rm cl}$-$r_v$-$Z$ point's closest \texttt{CMC} counterpart, we assign each of its BBH mergers the total weight
\begin{equation}
    \label{eq:totalweight}
    w_{\rm tot} = \frac{f_{\rm cl} f_{\rm bound} f_{\rm disrp} \psi(z_f)}{\braket{M_{\rm GC}}} \times w_{\rm form} \times w_{\rm scale}.
\end{equation}

For each cluster formed at redshift $z_f$, we propagate each BBH merger in the corresponding \texttt{CMC} model forward in cosmic time, weighted by Equation~\ref{eq:totalweight}. The cosmological merger lookback time is $\tau = t_L(z_f) - t_{\rm merge}$, where $t_L$ is the lookback time function, and $t_{\rm merge}$ is the simulation merger time after formation. Each BBH merger is shifted to the appropriate redshift bin $z_{\rm merge} = t_L^{-1}(\tau)$, and accumulated with weight $w_{\rm tot}$. Thus, $P$ in the convolution integral of Equation~\ref{eq:convolution} is approximated by the discrete delay-time distribution of each \texttt{CMC} model, and the total convolution is numerically evaluated by summation throughout formation redshifts $z_f \in [0,20]$, with step size of $0.1$.

The results of this process are shown in Figure~\ref{fig:rate}, which illustrates the contribution of different cluster types to the volumetric BBH merger rate over cosmic time. Our formation model yields a total BBH merger rate amplitude of $8.4~{\rm Gpc^{-3}\,yr^{-1}}$ at $z=0$, and a peak of $17.5~{\rm Gpc^{-3}\,yr^{-1}}$ at $z=1.2$, roughly consistent with previous studies \citep[e.g.,][]{RodriguezLoeb2018,AntoniniGieles2020mergers,Mapelli2022,Ye2024redshift}.

The top-left panel of Figure~\ref{fig:rate} shows the BBH merger rate for different primary mass groups as a function of redshift. Lower primary masses peak at lower redshifts, with peaks at $z=1.0$, $1.4$, and $1.8$ for primary masses of $M_1 < 20\,M_\odot$, $20\,M_\odot < M_1 < 40\,M_\odot$, and $40\,M_\odot < M_1$, respectively. This primarily occurs due to the higher star formation metallicity at lower redshifts (Equation~\ref{eq:muz}), which in turn reduces the average BH mass. Since most BBH mergers in a globular cluster occur within the first few Gyrs of its lifetime \citep[e.g.,][]{Rodriguez2016mergers,Kremer2020modeling}, the merger redshift remains closely tied to the cluster formation epoch. Clusters forming at lower redshifts are born with higher metallicity, producing lower-mass BHs and, consequently, lower-mass BBH mergers. Additionally, within individual clusters, higher-mass BBH systems tend to merge earlier in the cluster's lifetime, due to shorter inspiral times and mass segregation/hardening timescales (due to larger cross sections for dynamical encounters). Both of these effects shift the peak merger rate of lower-mass systems toward lower redshifts. Low-mass mergers with $M_1 < 20\,M_\odot$ dominate the merger rate at low redshift, composing 78\% of mergers at $z=0$, underlining the importance of high-metallicity clusters to the local-Universe merger rate.

The top-right panel of Figure~\ref{fig:rate} shows the contribution of different initial cluster masses to the overall merger rate, from $5 \leq \log(M_{\rm cl}/M_\odot) \leq 7$. Although a single high-mass cluster with $\log(M_{\rm cl}/M_\odot) > 6.5$ can contribute tens or hundreds of times more BBH mergers than a lower-mass cluster \citep[e.g.][]{Mai2026colossus,Bruel2025}, their contribution to the merger rate in our formation model is suppressed by the exponential Schechter term in Equation~\ref{eq:schechter}, since they lie beyond the fiducial Schechter mass $M_S = 10^{6.26}\,M_\odot$. Instead, clusters with initial mass around $\log(M_{\rm cl}/M_\odot) = 5.8$ contribute most to the merger rate at all redshifts. We explore how variations in the Schechter mass $M_S$ can change the overall merger rate amplitude and the contribution of high-mass clusters in Section~\ref{sec:schechter}. Additionally, as shown, the merger rates from higher-mass clusters peak at lower redshift than lower-mass clusters. This is due to the longer relaxation times of higher-mass clusters \citep[$t_{\rm rlx} \propto M_{\rm cl}^{1/2}$; e.g.][]{BinneyTremaine2008}, which lengthen their delay-time distributions and shift their merger peaks closer to present day.

The bottom-left panel of Figure~\ref{fig:rate} shows the contribution of different cluster metallicities to the merger rate over redshift, from $-2 \leq \log(Z/Z_\odot) \leq 0.5$. The range $-6 \leq \log(Z/Z_\odot) < -2$ is also included in the formation model, but is not visible due to its low amplitude. The gaps between various metallicity curves are due to the discrete metallicity binning process described previously. High-metallicity clusters dominate the rate at low redshift, with $Z/Z_\odot \geq 0.2$ clusters contributing 8.3 times as many mergers than $Z/Z_\odot < 0.2$ at $z=0$. This suggests that the majority of dynamical BBH mergers observed today are likely low-mass mergers from locally-formed high-metallicity clusters.

Finally, the bottom-right panel of Figure~\ref{fig:rate} shows the contribution of different cluster virial radii to the overall merger rate over redshift, from $0\,\mathrm{pc} \leq r_v \leq 4\,\mathrm{pc}$. The merger rate amplitudes of different $r_v$ values monotonically increase with decreasing $r_v$, due to increasing cluster density \citep[e.g.,][]{Kremer2020modeling}. Since delay-time distributions do not vary significantly with $r_v$ after initial BH subsystem formation, each of the $r_v$ curves have identical shape with respect to redshift, unlike cluster mass or metallicity.

\begin{figure}
    \includegraphics[width=\linewidth]{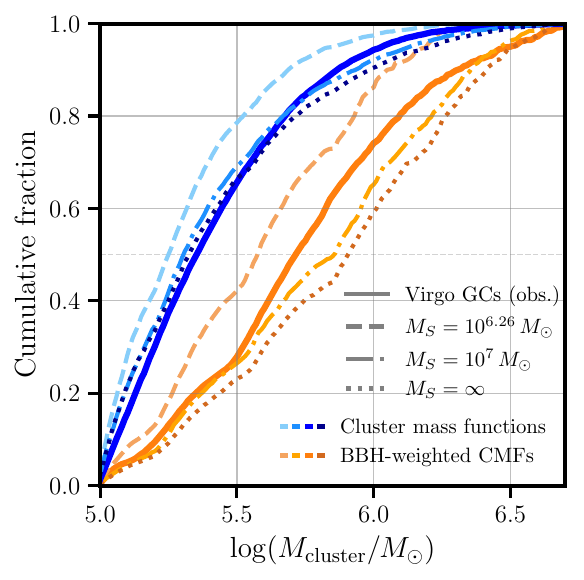}
    \caption{Cumulative present-day cluster mass functions (blue) for observed Virgo GCs, and for our analytic formation model with varying values of the Schechter mass. We find $M_S = 10^7\,M_\odot$ best matches the Virgo sample. We also show the same mass functions weighted by total number of BBH mergers (orange). Roughly independent of Schechter mass, half of all dynamical BBH mergers come from the heaviest $\sim15\%$ of dense clusters, whose cutoff mass increases with the chosen Schechter mass.}
    \label{fig:cmf}
\end{figure}

\section{Comparison with Virgo clusters}
\label{sec:virgo}

To assess the validity of our cluster formation model, we compare the present-day properties of our formation scheme to the present-day properties of observed GCs in the Virgo Supercluster. For Virgo GC properties, we use the data from the ACS Virgo Cluster Survey \citep{Jordan2009} to obtain present-day mass and half-light radii for about $1.3\times10^4$ Virgo clusters. To compare to these old GCs, we exclude all 39 \texttt{CMC} models which dissolved before $12~{\rm Gyr}$ from this section.

\subsection{Monte Carlo cluster sampling}

To obtain present-day properties of our formation model, we perform Monte Carlo sampling of globular clusters under the property distributions in Section~\ref{sec:formation}, and apply simple scaling arguments to estimate their present-day mass and half-light radii. Since the ACS Virgo Cluster Survey primarily observed old, early-type elliptical and lenticular galaxies, for this section we focus only on formation redshifts $z_f \geq 1.5$, corresponding to lookback times of $t_L \gtrsim 9.5\,\text{Gyr}$. Integrating Equation~\ref{eq:rgc} from $z_f=1.5$ to $20$, we obtain a number density of $2.29\,\text{Mpc}^{-3}$ for old globular clusters in the local Universe, consistent with empirical estimates which generally span the range $0.3-3\,{\rm Mpc}^{-3}$ depending on a variety of assumptions \citep[e.g.,][]{Phinney1991,PortegiesZwartMcMillan2000,Harris2013,Hudson2014,Rodriguez2015,Harris2017,AntoniniGieles2020mergers,Carlsten2022}.

\begin{figure*}
    \includegraphics[width=\textwidth]{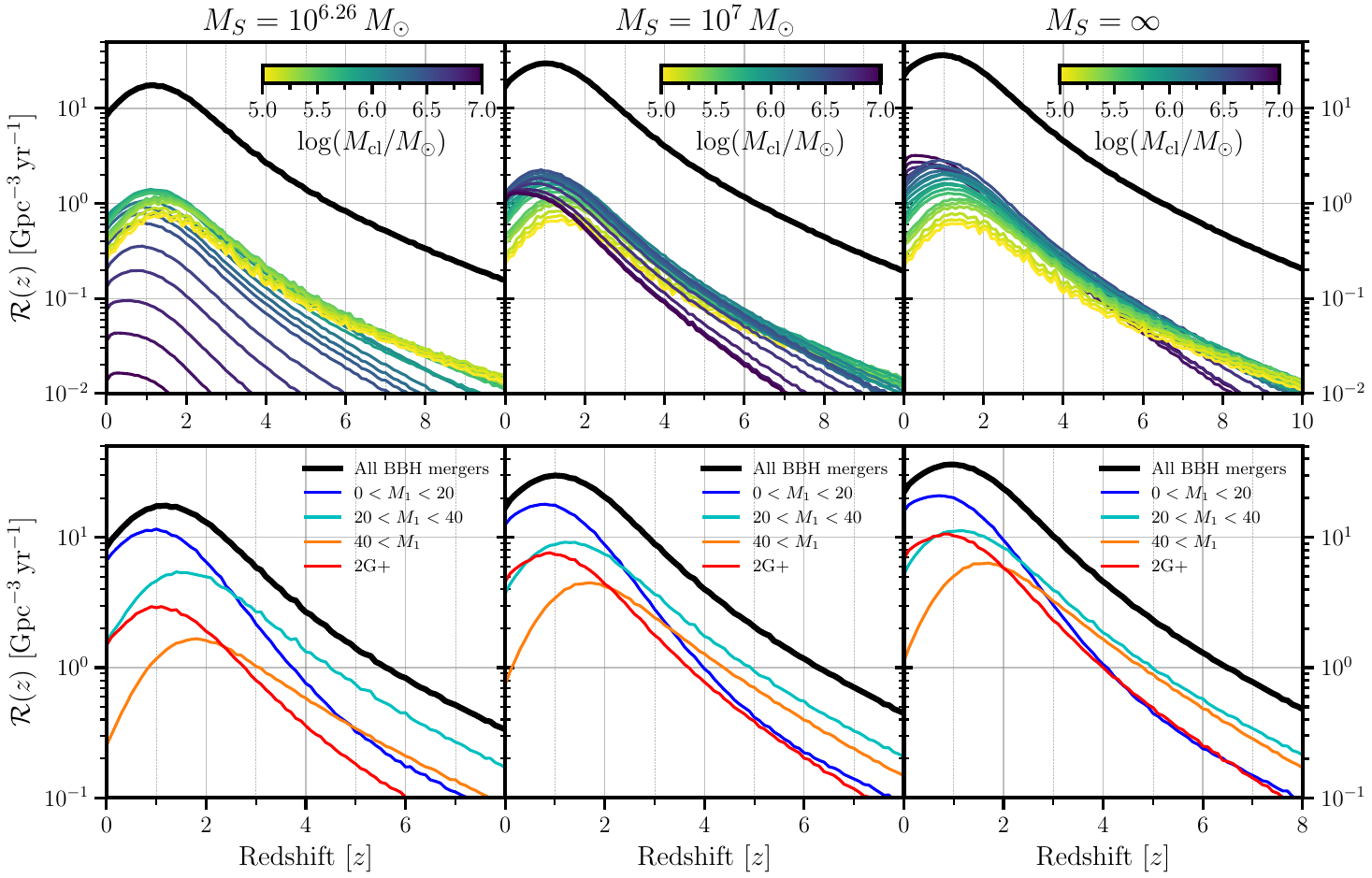}
    \caption{Top row: Contributions of different merger masses to the dynamical BBH merger rate over redshift (same as top right panel of Figure~\ref{fig:rate}), with varying Schechter mass values. As $M_S$ increases, the exponential Schechter suppression of high-mass clusters is released, allowing them to compose an increasing fraction of the total merger rate. The amplitude of $\mathcal{R}(z)$ can increase by up to threefold as a result of Schechter mass variation. Bottom row: merger rates for different primary mass ranges and hierarchical generation (same as top left panel of Figure~\ref{fig:rate}). As Schechter mass increases, high-mass clusters contribute more hierarchical mergers, raising the 2G+ and high-primary-mass rate amplitudes.}
    \label{fig:schechter_rate}
\end{figure*}

Under the property distributions described in Section~\ref{sec:formation}, we perform $10^5$ total random cluster draws, distributed amongst formation redshifts $z_f=1.5$ to $20$ in proportion to the redshift-dependent star formation rate density from Equation~\ref{eq:sfrd}. We map each of these random cluster draws to their closest \texttt{CMC} model counterpart using the same nearest-neighbor process described in Section~\ref{sec:formation}. Let $M_{\text{cl},0}$ and $r_v$ denote the initial mass and virial radius of a cluster formed at redshift $z_f$, and let $M_{\rm cl}^{\texttt{CMC}}(t)$ and $r_h^{\texttt{CMC}}(t)$ represent the time-dependent mass and $r_h$ curves of the cluster's nearest $\texttt{CMC}$ model. Then we estimate the cluster's present-day mass $M_{\text{cl},p}$ and half-light radius $r_{h,p}$ as
\begin{equation}
    \label{eq:presentscale}
    \begin{aligned}
        M_{\text{cl},p} & = M_{\text{cl},0} \times \frac{M_{\rm cl}^{\texttt{CMC}}(t=t_L(z_f))}{M_{\text{cl}}^{\texttt{CMC}}(t=0)}\\
        r_{h,p}  & = 0.6r_v \times \frac{r_h^{\texttt{CMC}}(t=t_L(z_f))}{r_{h}^{\texttt{CMC}}(t=0)}
    \end{aligned}
\end{equation}
where $0.6r_v \approx r_{h,0}$ is the approximate conversion from initial $r_v$ to initial $r_h$.

After performing this procedure for each random cluster draw, we compare the estimated present-day properties of our formation model to the properties of observed Virgo GCs. In Figure~\ref{fig:cmf}, we show cumulative present-day mass functions for our model alongside that of Virgo GCs in blue, and the mass function weighted by total number of BBH mergers in orange. The latter curve for Virgo clusters was obtained from the procedure in \cite{Mai2026colossus}, which utilizes a similar nearest-neighbor scheme but cannot account for merger delay-time distributions, as precise Virgo GC ages are unknown.

\subsection{Schechter term variation}
\label{sec:schechter}

Between the GC mass range of $\log(M_{\rm cl}/M_\odot) \in [5.0,7.0]$, our initial GC mass function with the Schechter mass of $M_S = 10^{6.26}\,M_\odot$ (Equation~\ref{eq:schechter}) roughly reproduces the shape of the Virgo distribution, but underproduces high-mass GCs. For comparison, in Figure~\ref{fig:cmf} we also plot present-day mass functions for two modified initial cluster mass functions: one with $M_S = 10^7\,M_\odot$, and one in the limit $M_S = \infty$, equivalent to $p(M) \propto M^{-2}$. We find that $M_S = 10^7\,M_\odot$ provides the best match to Virgo observations.

Within our cluster formation model, we find that Schechter variation does not significantly impact GC number density. In Equation~\ref{eq:rgc}, raising $M_S$ increases $\braket{M_{GC}}$ and $f_{\rm cl}$ about equally, and since we normalize based on a fixed star formation rate density, the integrated GC number density changes only slightly with $M_S$ variation. Integrating Equation~\ref{eq:rgc} from $z=1.5$ to $20$, at $M_S = 10^{6.26}\,M_\odot$, we previously found a GC number density of $2.29\,{\rm Mpc}^{-3}$; at $M_S = 10^{7}\,M_\odot$ and $\infty$, we find number densities of $2.28$ and $2.19\,{\rm Mpc}^{-3}$, respectively.

Comparing the BBH-weighted cluster mass functions of all three Schechter variations, we find that roughly independent of Schechter mass, the heaviest $\sim15\%$ of clusters are responsible for half of all dynamical BBH mergers. For Schechter masses of $M_S/M_\odot = 10^{6.26}$, $10^7$, and $\infty$, this corresponds to cluster masses beyond $\log(M_{\rm cl}/M_\odot) = 5.58$, $5.87$, and $5.91$, respectively.

As shown in Figure~\ref{fig:rate}, while a single high-mass cluster can contribute hundreds of times more mergers than lower-mass clusters, their contribution to the overall merger rate can be highly suppressed by the exponential Schechter term of the initial cluster mass function. Thus, the exact value of the Schechter mass $M_S$ can have a strong impact on the dynamical BBH merger rate, especially its exact amplitude and 2G+ fraction.

In Figure~\ref{fig:schechter_rate}, we show how the contribution of different cluster masses to the volumetric BBH merger rate changes with Schechter mass, using the same three values of the Schechter mass as the previous section: $M_S/M_\odot = 10^{6.26}$, $10^7$, and $\infty$. Under the fiducial $M_S$ value of $10^{6.26}\,M_\odot$, the merger rate from high-mass clusters is suppressed, and the merger rate is dominated by intermediate-mass clusters around $M_{\rm cl}\sim10^{6}\,M_\odot$. For $M_S=10^7\,M_\odot$, the highest contributions come from $M_{\rm cl}\sim10^{6.5}\,M_\odot$, and the highest-mass clusters at $M_{\rm cl}=10^7$ dominate the merger rate at $M_S = \infty$. The total merger rate amplitudes for all three Schechter variations are $\mathcal{R}(z) = 8.5$, $16.9$, and $22.1~{\rm Gpc^{-3}\,yr^{-1}}$ respectively, indicating up to a threefold increase in total dynamical merger rate as a result of Schechter variation.

In the bottom row of Figure~\ref{fig:schechter_rate}, for each Schechter variation, we also show merger rates for different primary mass groups and hierarchical generation. While the shapes of all these curves remain nearly identical between Schechter variations, their relative amplitudes differ. The high escape velocities of high-mass clusters aid in retention of second- and higher-generation BBH merger remnants, allowing them to dominate the 2G+ merger rate \citep[e.g.][]{Mai2026colossus}; thus, as Schechter mass increases and high-mass clusters become more common, the 2G+ fraction increases, alongside the heavier primary mass curves.

\begin{figure}
    \includegraphics[width=\linewidth]{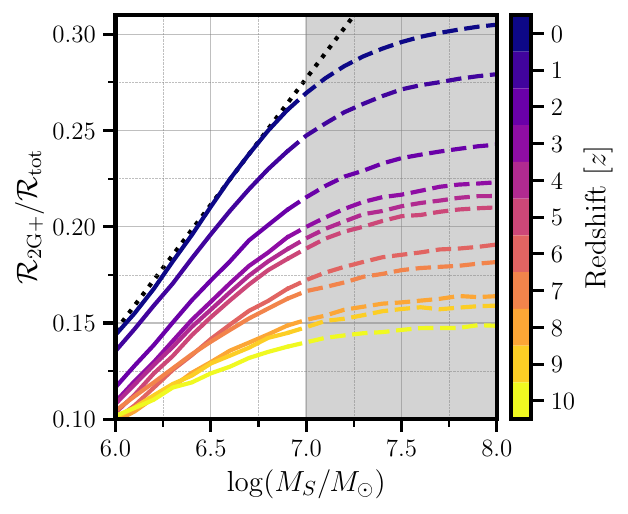}
    \caption{Fraction of 2G+ mergers in the total dynamical merger rate for different $z$, as a function of Schechter mass. The 2G+ fraction is approximately linear with $\log(M_S/M_\odot)$ until $M_S = 10^7\,M_\odot$. At high $z$, clusters have not lived long enough to create a substantial 2G BH population, so the 2G+ merger fraction is low. We shade the region past $\log(M_S/M_\odot) > 7$ to show the end of our model coverage in cluster mass. In dotted black, we show the linear fit for $z=0$ given in Equation~\ref{eq:2gfit}.}
    \label{fig:2gfrac}
\end{figure}

\begin{figure}
    \includegraphics[width=0.9\linewidth]{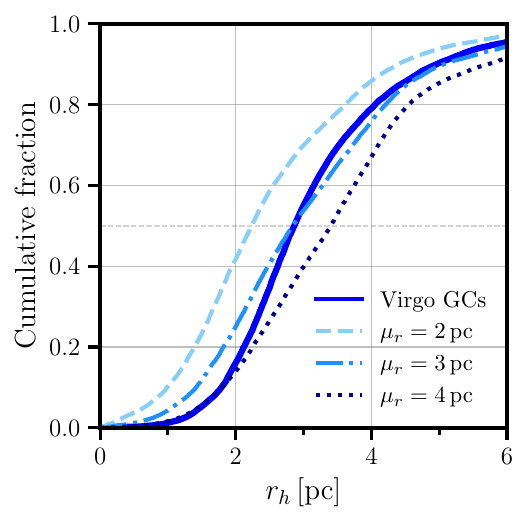}
    \caption{Comparison between cumulative present-day $r_h$ functions for observed Virgo GCs, and for our analytic formation model with varying values of $\mu_r$. We find that a value of $\mu_r = 3\,{\rm pc}$ best matches the Virgo $r_h$ distribution.}
    \label{fig:rhf}
\end{figure}

\begin{figure*}
    \includegraphics[width=\textwidth]{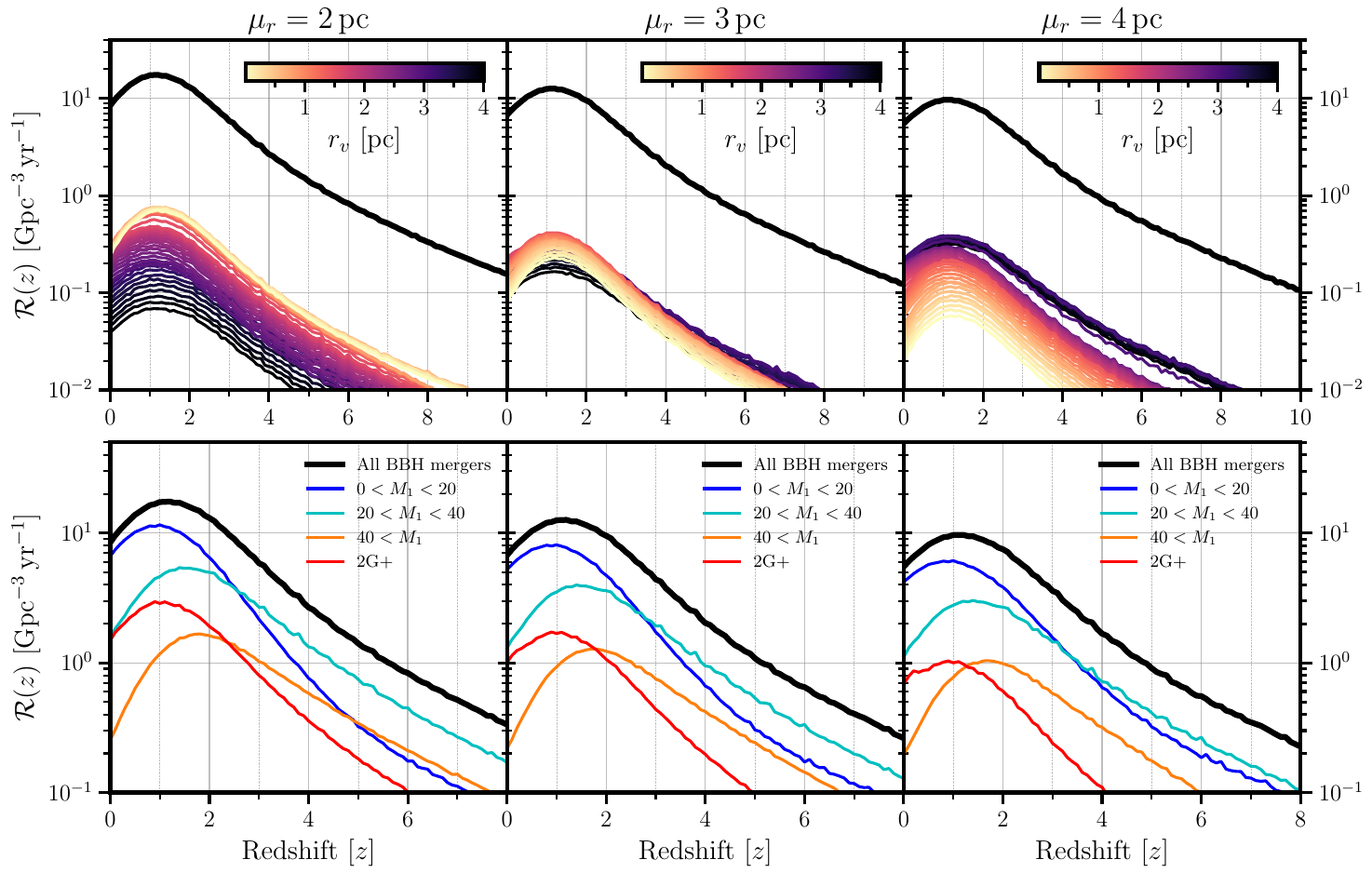}
    \caption{Top row: contributions of initial cluster virial radii to the dynamical BBH merger rate over redshift (same as bottom right panel of Figure~\ref{fig:rate}), with varying $\mu_r$ (Equation~\ref{eq:rvpdf}). As $\mu_r$ increases from $2\,{\rm pc}$, the contribution of the densest clusters become suppressed, and the puffiest clusters eventually dominate the merger rate due to their prevalence in number density. Bottom row: merger rate curves split by primary mass and hierarchical generation (same as top left panel of Figure~\ref{fig:rate}). As $
    \mu_r$ increases, the 2G+ rate is correspondingly reduced due to lower escape velocities in puffier clusters.}
    \label{fig:mur_rate}
\end{figure*}

For $M_S/M_\odot = 10^{6.26}$, $10^7$, and $\infty$ we find the fraction of 2G+ mergers at $z=0$ to be $17.6\%$, $26.9\%$, and $30.9\%$, respectively. Between $M_S/M_\odot=10^6$ and $10^7$, we find that the 2G+ fraction at $z=0$ closely follows a linear relationship with $\log(M_S/M_\odot)$:
\begin{equation}
    \left.\frac{\mathcal{R}_{\rm 2G+}(z)}{\mathcal{R}_{\rm tot}(z)}\right|_{z=0} = (0.131 \pm 0.002)\log\left(\frac{M_S}{M_\odot}\right)-(0.64\pm0.01)
    \label{eq:2gfit}
\end{equation}
which asymptotes to $30.9\%$ as $M_S$ tends to infinity. However, we note that our formation model is restricted to cluster masses below $10^7\,M_\odot$, so we do not physically interpret Schechter mass values above $10^7\,M_\odot$. The asymptote to $30.9\%$ is simply the total 2G+ fraction of our models in the limit $p(M) \propto M^{-2}$.

In Figure~\ref{fig:2gfrac}, we plot the fraction of 2G+ mergers versus Schechter mass for $z=0$ through $10$. For all redshifts, the 2G+ fraction follows approximately a linear relationship with $\log(M_S/M_\odot)$ which transitions to an asymptote after $M_S = 10^7\,M_\odot$. For all values of $M_S$, 2G+ fraction decreases at higher redshift. This is simply because at high $z$, clusters have not had enough time to form a substantial 2G black hole population, so the 2G+ merger fraction remains low.

\subsection{Cluster radius variation}

In a similar manner to cluster mass distributions, we also explore how variations in initial virial radii ($r_v$) affect present-day half-light radii ($r_h$) distributions and BBH merger populations. We vary the mean initial virial radius $\mu_r$ in Equation~\ref{eq:rvpdf}, and fix $M_S = 10^{6.26}\,M_\odot$ for this section. In Figure~\ref{fig:rhf} we compare cumulative present-day $r_h$ distribution between Virgo GCs and our cluster formation model; we find that between $\mu_r = 2$, $3$, and $4\,\rm{pc}$, $\mu_r = 3\,\rm{pc}$ provides the best match with Virgo observations, producing a median present-day $r_h$ value of $2.92\,\rm{pc}$, compared to Virgo's $2.87\,\rm{pc}$.

We show the impact of $\mu_r$ variation on the BBH merger rate in Figure~\ref{fig:mur_rate}. In the top row, we show the contribution of different initial cluster $r_v$ to the BBH merger rate over redshift, for varying $\mu_r$. Since denser clusters produce higher BBH merger yields \citep[e.g.,][]{Kremer2020modeling}, at $\mu_r = 2\,{\rm pc}$ where high-$r_v$ and low-$r_v$ clusters form approximately equally, the densest low-$r_v$ clusters contribute most to the overall merger rate. At $\mu_r = 3\,{\rm pc}$, the lower yields of high-$r_v$ clusters are balanced by their increased prevalence, and all $r_v$ values contribute approximately equally to the merger rate. Finally, at $\mu_r = 4\,{\rm pc}$, the contribution trend at $\mu_r = 2\,{\rm pc}$ is fully reversed, and the merger rate is instead dominated by puffy high-$r_v$ clusters, which compose the majority of clusters at this $\mu_r$ value. For these three values of $\mu_r$, the merger rate amplitudes at $z=0$ are $8.4$, $6.7$, and $5.5~{\rm Gpc^{-3}\,yr^{-1}}$ respectively, which indicate at most a $1/3$ reduction in merger rate amplitude as a result of $\mu_r$ variation. Compared to $M_S$ variation, which can increase the amplitude by up to a factor of $3$, $\mu_r$ variation is a relatively minor effect.

\begin{figure}
    \includegraphics[width=\linewidth]{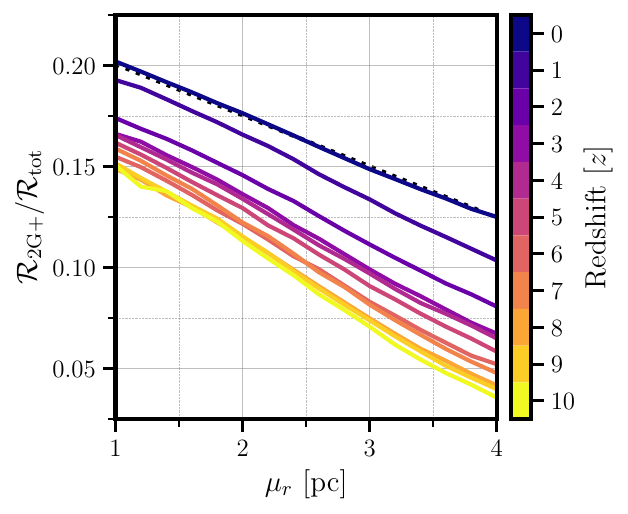}
    \caption{Fraction of 2G+ mergers in the total dynamical merger rate for different $z$, as a function of mean virial radius $\mu_r$. For all redshifts, the 2G+ fraction decreases linearly with $\mu_r$. In dotted black, we show the linear fit to $z=0$ given in Equation~\ref{eq:2gmur_fit}.}
    \label{fig:2gfrac_mur}
\end{figure}

Since higher-radius clusters have a lower escape velocity, puffier clusters have a reduced yield of hierarchical mergers, similar in effect to lowering cluster mass. In the bottom row of Figure~\ref{fig:mur_rate}, we again show merger rates for different primary masses and hierarchical generation. The 2G+ fractions for $\mu_r = 2$, $3$, and $4\,{\rm pc}$ are $17.6\%$, $14.9\%$, and $12.5\%$, respectively.

In Figure~\ref{fig:2gfrac_mur}, we show the fraction of 2G+ mergers versus $\mu_r$ for $z=0$ through $10$. For the same reasons discussed with Figure~\ref{fig:2gfrac}, 2G+ fraction monotonically decreases with redshift. For all $z$, the 2G+ fraction decreases approximately linearly with $\mu_r$. At $z=0$, the fraction of 2G+ mergers closely follows the linear relationship
\begin{equation}
    \left.\frac{\mathcal{R}_{\rm 2G+}(z)}{\mathcal{R}_{\rm tot}(z)}\right|_{z=0} = (-0.025\pm0.001)\mu_r+(0.226\pm0.001).
    \label{eq:2gmur_fit}
\end{equation}

\section{Black Hole Merger Properties}
\label{sec:bbh}

Throughout the cosmic integration process in Section~\ref{sec:integration}, we can also individually track volumetric merger rates for each of the $\sim 20{,}000$ BBH mergers in our combined model catalog, allowing us to create detailed property distributions of dynamical BBH mergers across all redshift. In this section, we specifically focus on mass distributions, and reserve discussion of other properties like spin and eccentricity evolution for future studies. We include spin and eccentricity information in our public data release. Unless otherwise specified, we fix $M_S = 10^{6.26}\,M_\odot$ and $\mu_r = 2\,{\rm pc}$ for this section, as in Figure~\ref{fig:rate}.

\begin{figure}
    \includegraphics[width=\linewidth]{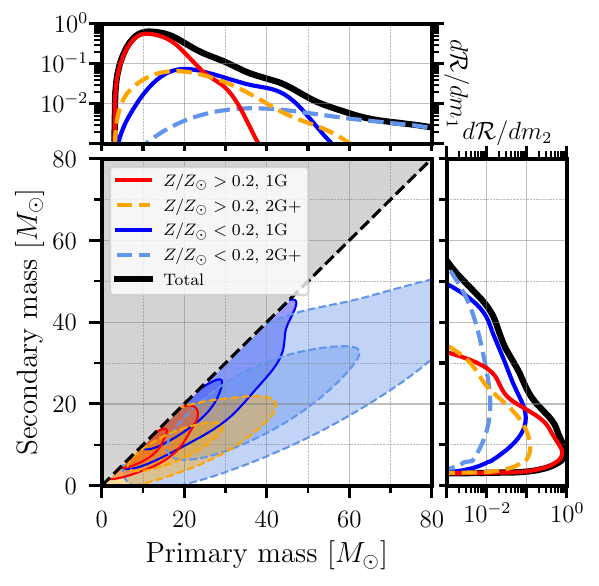}
    \caption{Primary versus secondary mass contours (50\%, 90\%) for dynamical BBH mergers at $z=0.2$, split by host cluster metallicity and hierarchical merger generation. The BH mass function of high-metallicity clusters concentrates their BBH mergers towards lower masses. Low-metallicity clusters have a much broader merger mass range, but occur at low rates at $z=0$.}
    \label{fig:c_m1m2}
\end{figure}

\begin{figure}
    \includegraphics[width=\linewidth]{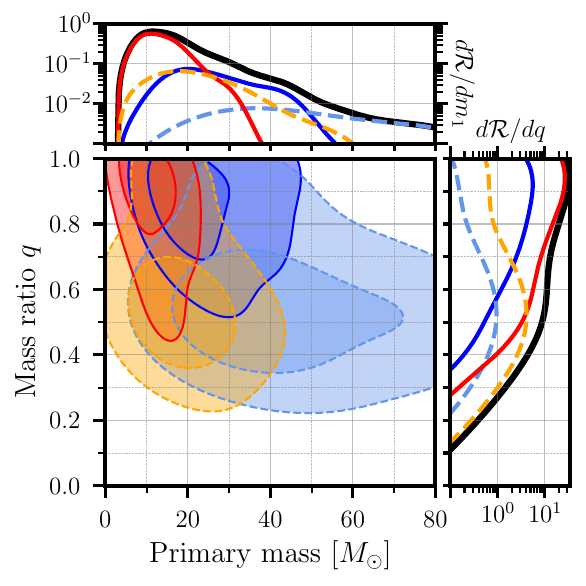}
    \caption{Mass ratio versus primary mass contours (50\%, 90\%) for dynamical BBH mergers at $z=0.2$, using the same contour colors as Figure~\ref{fig:c_m1m2}. The mass ratio distributions are shaped nearly identically between low- and high-metallicity mergers, differing only in their amplitudes.}
    \label{fig:c_mratio}
\end{figure}

\begin{figure*}
    \includegraphics[width=\textwidth]{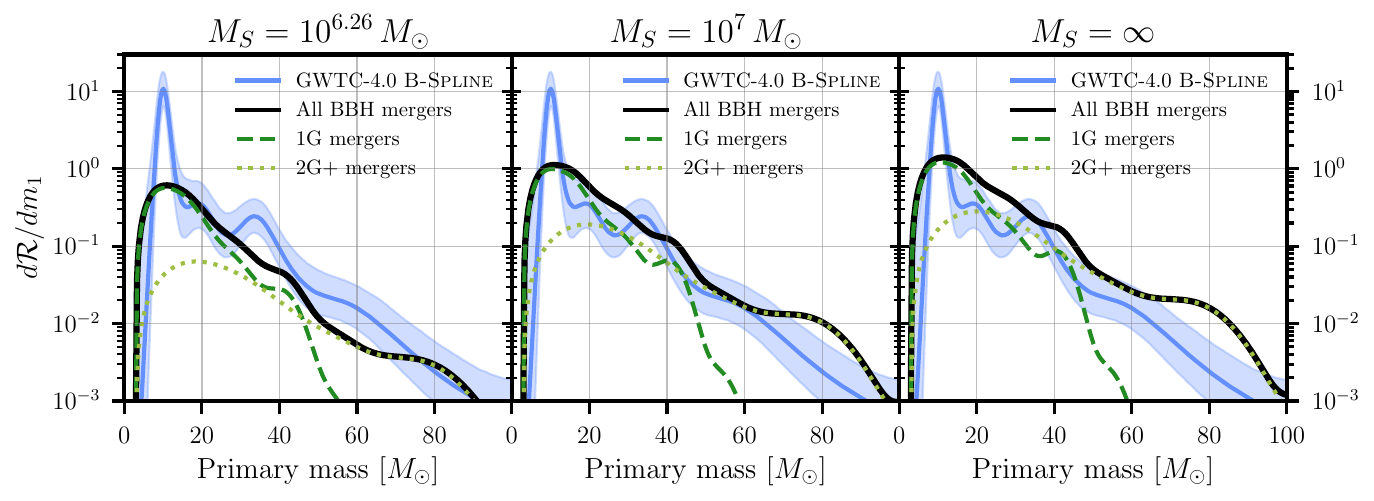}
    \caption{$d\mathcal{R}/dm_1$ distributions at $z=0.2$ for three variations of the Schechter mass, compared with the \textsc{B-Spline} model from GWTC-4.0. We apply smoothing to our rate curves to capture global trends independent of our particular BH mass function. The overall shape of the $d\mathcal{R}/dm_1$ distributions remain consistent when increasing $M_S$, except for a stronger tertiary peak at $80\,M_\odot$ due to increasing 2G+ fraction at high $M_S$. Several features of our mass distributions are qualitatively comparable to the \textsc{B-Spline} model: a low-mass peak at $10\,M_\odot$, a secondary peak at $40\,M_\odot$ due to our implemented pair-instability cutoff, and an extended high-mass tail due to hierarchical mergers. Overall, $M_S=10^7\,M_\odot$ has the best-matching normalization to the \textsc{B-Spline} rate.}
    \label{fig:bspline}
\end{figure*}

In Figure~\ref{fig:c_m1m2}, we show $50\%$ and $90\%$ contours for dynamical BBH mergers at $z=0.2$ in $m_1$-$m_2$ space, split by hierarchical generation and host cluster metallicity. We apply smoothing to these contours and their projections, in order to isolate high-level trends in redshift evolution from the granular details of our particular BH mass function (see Section~\ref{sec:cmc}). The qualitative metallicity trend in Figure~\ref{fig:m1m2-crb} can also be observed here: as a result of their bottom-heavy BH mass functions, high-metallicity clusters preferentially produce BBH mergers with primary mass $<20\,M_\odot$, while 1G mergers in low-metallicity clusters can reach up to $\sim40\,M_\odot$. Hierarchical mergers extend the primary mass range of both metallicity types to roughly double that of their 1G counterparts: 2G+ high-$Z$ mergers up to $\sim 40\,M_\odot$, and low-$Z$ mergers beyond $80\,M_\odot$.

In Figure~\ref{fig:c_m1m2}, we also show the rate density curves $d\mathcal{R}/dm_1$ and $d\mathcal{R}/dm_2$, also split by host cluster metallicity and hierarchical generation. Several broad features can be observed in the $d\mathcal{R}/dm_1$ distribution: a low-mass peak at about $10\,M_\odot$ attributed to 1G mergers from high-metallicity clusters, a slight plateau at $40\,M_\odot$ due to \texttt{CMC}'s pair-instability cutoff of $40.5\,M_\odot$, and an high-mass tail between $50\,M_\odot<m_1<100\,M_\odot$ dominated by hierarchical mergers from low-metallicity clusters.

While not visible in Figure~\ref{fig:c_m1m2}, past $100\,M_\odot$, the low-$Z$ hierarchical merger rate falls off, and the rate is instead dominated by both 1G and 2G+ mergers from high-$Z$ clusters. Since high-mass BHs have shorter mass segregation and inspiral timescales, they merge very early on in the cluster's lifetime; thus, high-mass BHs from low-$Z$ clusters have already merged at high redshift and are not observable, but the continuous stream of recently formed high-$Z$ clusters can supply high-mass BBH mergers in the local Universe. These high-mass 1G BHs are formed through repeated stellar mergers in the first few Myr of the cluster's lifetime \citep[e.g.,][]{Gonzalez2021,Ye2026}.

In Figure~\ref{fig:c_mratio}, we show the same merger contours at Figure~\ref{fig:c_m1m2} but instead in $m_1$-$q$ space, where $q$ is the mass ratio $m_2/m_1$. We also show the rate density curve $d\mathcal{R}/dq$. Due to mass segregation, BBH systems preferentially form with equal-mass companions, so the primary peak of $d\mathcal{R}/dq$ lies close to $q\simeq1$. There is also a secondary peak at $q\simeq0.5$ caused by 2G+1G hierarchical mergers, which on average have a mass ratio of $0.5$. In $q$ space, high-metallicity and low-metallicity clusters have nearly identical distributions, but occur at different amplitudes at $z=0$.

In Figure~\ref{fig:bspline}, we provide a direct comparison of our $d\mathcal{R}/dm_1$ curves with the \textsc{B-Spline} model of GWTC-4.0 \citep{GWTC40}, for the three $M_S$ values discussed previously. We split the $d\mathcal{R}/dm_1$ curves of our formation model by hierarchical generation, combining both low- and high-metallicity mergers. Several features in our distributions are qualitatively comparable to those of the \textsc{B-Spline} distribution: a peak at $10\,M_\odot$, a slight change in slope around $35\,M_\odot$, and an extended high-mass tail. This shape remains consistent between $M_S$ values except for a higher 2G+ fraction at higher $M_S$, which causes a stronger tertiary peak at around $80\,M_\odot$. Although smoothed over in the overall $d\mathcal{R}/dm_1$ curve, we are able to qualitatively reproduce the $20\,M_\odot$ \textsc{B-Spline} sub-peak through 2G+ mergers. In our model, we attribute the primaries of these mergers to the 2G remnants of the $10\,M_\odot$ peak. Assuming all of the observed mergers in the $20\,M_\odot$ sub-peak are indeed 2G mergers from high-metallicity clusters, we find that $M_S \approx 10^7$ provides the best normalization to this peak, as well as the rest of the \textsc{B-Spline} distribution.

\section{Summary and Conclusions}
\label{sec:conclusion}

\subsection{Summary}

In this study, we have used an analytic cluster formation model to study dynamical BBH merger rates and properties over cosmic time, and how these results change when varying the initial property distributions of dense star clusters. We specifically focus on the contribution of high-mass clusters to the BBH merger rate, which lacked model coverage in similar previous studies. We identify the following key results:

\begin{enumerate}
    \item We obtain merger rate amplitudes for dynamical BBH mergers across all redshift, finding the volumetric rate at $z=0$ to be $8.4{-}22.1~{\rm Gpc^{-3}\,yr^{-1}}$, depending on assumptions about the high-mass tail of the cluster mass function. We find that in the local Universe, the dynamical BBH merger rate is dominated by low-mass ($m_1<20\,M_\odot$) mergers from recently-formed, high-metallicity ($\log Z/Z_\odot>-1$) clusters, consistent with previous studies \citep[e.g.,][]{Ye2026}.
    \item The contribution of high-mass clusters to the overall merger rate is highly sensitive to the assumed Schechter mass in the initial cluster mass function, $p(M) \propto (M/M_S)^{-2}\exp(-M/M_S)$. We find that such Schechter mass variation can increase the local merger rate amplitude up to a factor of $\sim3$ from the canonical value of $M_S = 10^{6.26}\,M_\odot$, and that a value of $M_S = 10^7\,M_\odot$ best matches the present-day mass distribution of Virgo globular clusters.
    \item Since high-mass clusters are the primary sources of hierarchical mergers, the local 2G+ fraction of dynamical mergers changes significantly with the assumed Schechter mass, ranging from $18\%$ for $M_S = 10^{6.26}\,M_\odot$ up to $31\%$ in the limit $M_S \to \infty$. Between $M_S=10^6$ and $10^7\,M_\odot$, the 2G+ fraction closely follows a linear relationship with $M_S$; future observational constraints on the dynamical 2G+ fraction may therefore provide an independent method for determining $M_S$.
    \item Variations in the initial virial radius $r_v$ distribution have a relatively small effect on BBH rate amplitudes compared to Schechter mass variations. We find that a value of $\mu_r = 3\,{\rm pc}$ best matches the Virgo half-light radius distribution, but that even in the pessimistic case of $\mu_r = 4\,{\rm pc}$, the overall BBH rate amplitudes and 2G+ fraction both decrease by at most 1/3 from the typically-assumed $\mu_r = 2\,{\rm pc}$.
    \item By normalizing our cluster formation history to the \citet{MadauFragos2017} star formation rate density, we find our bottom-line merger rate amplitudes and cluster number density estimates remain robust to variations in initial cluster property distributions, varying at most by factors of order unity. In particular, our GC number density estimates ($\sim2.3\,{\rm Mpc}^{-3}$) are remarkably close to observed values without fine-tuning, for all values of $M_S$.
\end{enumerate}

We release the volumetric rates and relative astrophysical weights calculated in this study for each of the $\sim20{,}000$ BBH mergers in the extended \texttt{CMC Cluster Catalog}, which can be used to explore the mass distributions discussed here, as well as the evolution of properties like merger spin and eccentricity over redshift.

\subsection{Discussion}

This analytic formation model can be used to rapidly assess the BBH merger populations produced by any cluster formation history, BH mass function, and cluster property distribution. We have focused here on high-level trends in merger metallicity over redshift, rather than the granular details of any particular BH mass function, but this formation model can indeed be used to assess the detailed $d\mathcal{R}$ curves produced by any assumed cluster formation history and BH mass function.

Here we assumed that cluster formation follows the cosmic star formation rate. In this case, cluster formation peaks at redshifts of $z\approx2$. However, proposed alternate scenarios include cluster formation caused by collapse of dark matter halos around the epoch of reionization, peaking around $6\lesssim z \lesssim12$ \citep[e.g.][]{Forbes2015,Trenti2015,RomeroShaw2021}. In this case, the contribution of high-metallicity clusters to the local merger rate, as well as the overall rate amplitude, may be significantly reduced compared to the former case.

We also assume a fixed value of $f_{\rm disrp} = 1/5$ for all redshifts and cluster masses, but detailed prescriptions of cluster dissolution timescales can be used to introduce a time- and cluster-property-dependent $f_{\rm disrp}$, of the form $f_{\rm disrp} \propto \exp[-t_{\rm merge}/t_{\rm disrp}(M_{\rm cl},r_v,...)]$. This can taper off the contribution of dissolving clusters gradually over time, instead of removing it entirely at formation. Dissolution is dominated by separate physical processes in different regions of $M_{\rm cl}$-$r_h$ space \citep[dynamical friction for the highest-mass clusters, tidal shocking for puffy clusters, evaporation for a typical cluster; e.g.][]{GnedinOstriker1997,BinneyTremaine2008}, but it is also strongly dependent on environmental factors like orbital galactocentric radius, galaxy type, and more \citep[e.g.][]{Boutloukos2003,Gieles2008}. Because of this environmental dependence, a detailed treatment of cluster dissolution is difficult to implement in an analytic cluster formation model such as this one. For discussion of these points in the context of BBH mergers, see e.g., \citet{FragioneKocsis2018,Rodriguez2020_SSC,AntoniniGieles2020mergers,Rodriguez2023}.\linebreak

\section*{Data Availability}

The list of all BBH mergers in our complete set of \texttt{CMC} models, along with volumetric rates for each merger calculated for three values of the Schechter mass ($10^{6.26}\,M_\odot$, $10^7\,M_\odot$, $\infty$), is available on Zenodo (DOI: \href{https://doi.org/10.5281/zenodo.23355226}{10.5281/zenodo.23355225}). 

\section*{Acknowledgments}

We thank Carl Rodriguez, Tom Wu, and Claire Ye for helpful discussions in preparation of this manuscript. K.K. acknowledges support from the UC San Diego Hellman Fellowship. F.K.\ acknowledges support from CIERA Postdoctoral Fellowship. 
This work used computing resources at the Quest high-performance computing facility provided by CIERA under NSF Grant PHY-2406802.
Quest is jointly supported by Northwestern University's Office of the Provost, the Office for Research, and Northwestern University Information Technology.

\bibliographystyle{aasjournal}
\bibliography{mybib.bib}

\end{document}